\documentclass[11pt,twoside]{article}

\usepackage{asp2004}
\usepackage{epsf}
\usepackage{psfig}
\usepackage{lscape}

\begin{document}
\title{Blazars and the emerging AGN/black hole X-ray binary
paradigm}   
\author{Philip Uttley}   
\affil{X-ray Astrophysics Laboratory, NASA Goddard Spaceflight Center,
Greenbelt, MD 20771}    

\begin{abstract} 
We briefly review the emerging paradigm which links the
radio-quiet and radio-loud classes of AGN to the different accretion
states observed in stellar mass black hole X-ray binary systems (BHXRBs), 
and discuss
the relevance of the AGN/BHXRB connection to blazar variability.
\end{abstract}
\vspace{-0.5cm}


\section{Introduction}
In recent years, a new paradigm has emerged which seeks to link AGN
behaviour with that of the stellar mass black hole X-ray binary systems.
There are currently two main approaches to this endeavour.  One is to
compare the temporal variability of AGN and BHXRBs.  This approach
has already yielded very interesting results, proving that characteristic
time-scales scale roughly linearly with black hole mass, and
linking the temporal variability
of several Seyfert galaxies to that of the BHXRB Cyg~X-1 in its high/soft
state (e.g. M$^{c}$Hardy et al. 2004 and
these proceedings).  The other approach
is based on the presence of jets in AGN and BHXRBs, and the
relation of the associated radio emission to the X-rays \citep{mer03,fal04}.
Also, the relationship of the
jet (its presence or absence, and the presence of jet
ejections) to the `state' of BHXRB systems has led to efforts
to describe the radio-loud/radio-quiet dichotomy in AGN in terms
of analogous states to those seen in BHXRBs (e.g. \citealt{mei01,fal04}).  

\section{BHXRB Accretion States}
The range of BHXRB behaviour is very diverse, but the following
broad picture has emerged,
primarily from studies of X-ray transients which move through
the full range of states during a single outburst (see \citealt{mcc05,fen05}
for reviews). \\ 
The {\it low/hard state} appears to correspond generally to low accretion rates
(less than a few per cent of Eddington),
with relatively low luminosities and X-ray spectra dominated by a hard 
(photon index $\Gamma<2$)
power-law continuum, with little or no obvious disk blackbody
emission (which can be
seen in the X-ray band in BHXRBs, due to their low black hole
masses and hence high disk temperatures).  Interestingly, all
low/hard state BHXRBs show the presence of radio jets, which become
stronger relative to the X-ray emission as the source drops to even lower
accretion rates, with the power output possibly becoming jet-dominated
at some point \citep{fen05}.\\
At higher accretion rates (few per cent Eddington or greater),
the sources can transition 
to a {\it high/soft state} where the emission is dominated by blackbody 
emission from the accretion disk, with only a very weak, steep ($\Gamma>2$)
power-law component present. The high/soft state is very interesting from
the point of view of understanding BHXRB jets, because they become much weaker
or may even disappear completely
during this state, with the radio
emission becoming undetectable ($>30$~times fainter than the low/hard state)\\
Bridging the power-law dominated and disk dominated states described above
is the {\it intermediate state} or {\it very high state}.  The different names
reflect the fact that the transition between a power-law-dominated and
disk-dominated state can in fact be observed at either relatively low
(few per cent Eddington) or rather high (tens of per cent Eddington)
accretion rates, but the behavior is similar in either case and both types
of transition state are now commonly thought of as being the same thing,
which we will refer to as the intermediate state.
As the name suggests, the intermediate state shows a spectrum intermediate 
between the low/hard and high/soft states, with relatively strong disk and
strong steep ($\Gamma>2$)
power-law contributions.  The fraction of luminosity contributed
by each component varies, so the state is in some sense
loosely defined and can be quite unstable, flipping between
more power-law dominated and disk dominated spectra
on relatively short time-scales.
The jet behaviour is also very interesting, with
powerful relativistic ejections occuring as the disk component
gets stronger (possibly connected to the quenching or
disappearance of the jet in the high/soft state) although a persistent jet 
can exist in the more power-law
dominated types of intermediate state.
\begin{figure}
\begin{center}
{\epsfxsize 0.9\hsize
 \leavevmode
 \epsffile{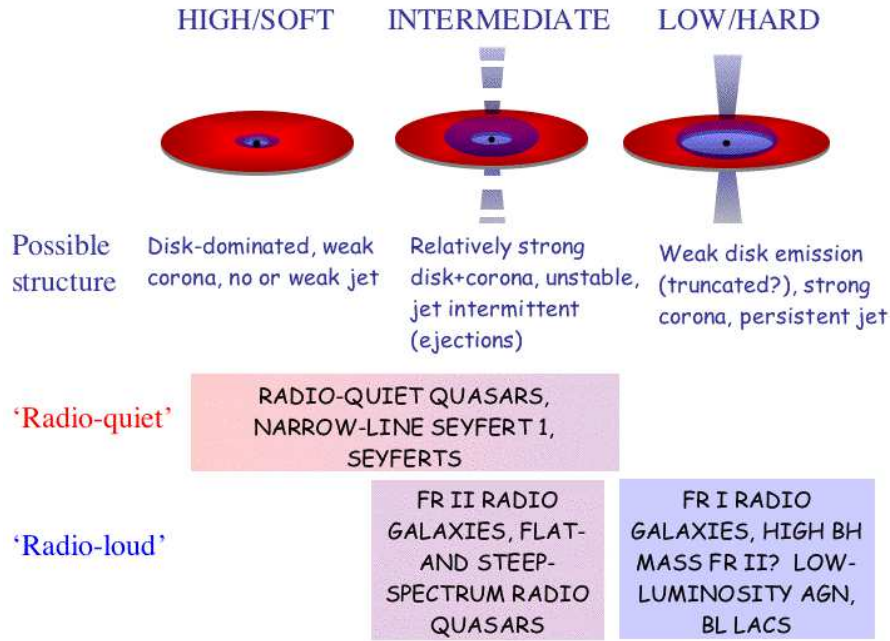}
}\caption{Black hole `Grand Unification'. The different classes
of AGN are grouped in terms of the accretion state they may occupy (see
\citealt{mei01,fal04} for a similar delineation).
} \label{grandunif}
\vspace{-1.0cm}
\end{center}
\end{figure}
\section{Black hole Grand Unification}
The existence of the same states in AGN would be highly interesting, not only
from a spectral and variability
point of view, but also in terms of understanding
the radio-loud/radio-quiet dichotomy,
because it is clear from BHXRB studies that states with and without
jets can occur in the same source.  Therefore, it is possible that the
presence of powerful
jets in AGN may be related to the accretion state, and not
necessarily to the black hole spin or AGN environment.  The types of state
seen in BHXRBs seem to be related to the interplay of the disk and power-law
emission (perhaps the latter from a corona), and the presence and/or strength
of the jet seems to be correlated with the presence of the power-law component.
With this simple picture, we can construct a picture of
`Black Hole Grand Unification', shown in Fig.~\ref{grandunif}, where the
distinction between radio-loud and radio-quiet AGN is governed by whether
the source occupies the strong-jet, strong-corona states (low/hard and 
parts of the unstable intermediate state) or the weak-jet, disk-dominated 
states (high/soft and disk-dominated parts of the intermediate state). \\
Note that we have classified low-luminosity AGN as low/hard state sources, as
previously suggested by \citep{ho05} in line with their low accretion rate
and their unusual spectral energy distributions (SEDs), which do not
show `big blue bumps' (suggesting that the disk is truncated), and
relatively strong radio emission (suggesting relatively strong jets). \\
The dividing line between the distinctive high/soft and low/hard states and
the intermediate state is murky. 
In BHXRBs these states can be distinguished from the intermediate
state by their relative stability and X-ray colours (since the relative
contributions of disk and power-law emission can be assessed in this way), 
but in AGN we cannot cleanly see most of the disk emission, 
which occurs in the FUV/EUV, so it is hard
to judge whether a source SED is disk or power-law dominated.  Changes
from power-law to disk dominated and back again can take hours to weeks
in BHXRBs, corresponding to thousands to millions of years for a 
$10^{8}$~M$_{\odot}$ black hole, so for any given AGN
we can only realistically observe a `snapshot' of it in a single state.
These limitations would make it difficult to distinguish between high/soft and
intermediate states with strong disk emission, hence we might expect
Seyferts and radio-quiet quasars to occupy either state.  \\
The 
intermediate and high/soft states are not tightly tied to
accretion rate, but are observed above a few per cent Eddington, whereas only
the low/hard state is seen at lower accretion rates \citep{mac03a}. 
Thus it
may be possible to distinguish low/hard and power-law dominated intermediate
states in the radio-loud sources, in terms of the 
luminosity and power of the source.  In general, one might expect
FR~II radio galaxies to correspond to the more powerful, higher accretion rate
intermediate states, while FR~I sources correspond to the low/hard state,
but more massive black holes may possess sufficiently
powerful jets that they correspond to FR~II sources in the low/hard state
also.

\section{Implications for Blazars}
Following the standard paradigm for unifying blazars with non-beamed
radio-loud AGN, we can roughly map the BL Lac sources on to the low/hard
state, and more powerful flat spectrum and steep spectrum
radio quasars on to the intermediate state.  The lack of 
significant optical permitted line emission, which distinguishes BL Lac
objects and maps them on to FR~I radio galaxies in the standard unification
paradigm may be related to the lack of any significant disk emission
which can drive the optical line emission (see \citealt{ho05}). \\
Interestingly, the timing behaviour of the different BHXRB
states may shed light
on variability behaviour of blazars, and offer a tantalising glimpse of
what {\it GLAST} might reveal in sufficiently long blazar light curves.
In BHXRBs,
strong quasi-periodic oscillations (QPOs), typically at around 1-10~Hz,
are often observed in the intermediate state, especially the power-law
dominated part, i.e. when a strong jet is present.  If the variations
are produced in an accretion flow which is coupled to the jet, similar 
behaviour might be seen in the blazars which occupy this state, but on 
time-scales of years (assuming linear scaling of QPO
time-scales with black hole mass).  In fact, periodic (or possibly
quasi-periodic) continuum variations on time-scales of years have been claimed
in a number of blazars, most notably OJ~287
(e.g. \citealt{val00}).  These candidate periodicities
\footnote{The sampling and number of cycles observed may
not yet be sufficient to show that 
the variations are strictly or quasi-periodic and 
not simply due to red noise (e.g. see discussion in
\citealt{vau05}).} are often interpreted
as being due to precession of the jet caused by a binary black hole system,
but it is interesting that they occur on similar relative time-scales to the
strong QPOs observed in the possibly analogous BHXRBs in the intermediate
state, where the QPOs are probably intrinsic to the accretion flow and
are not attributed to binary motion.  When {\it GLAST} is launched, it will
provide high quality gamma-ray light curves for many blazars, which, if the
mission is flown for sufficient duration, may be sufficient to detect
periodicities in the lower-mass blazars and test the remarkable AGN/BHXRB 
connection still further.

\end{document}